\documentclass[journal]{IEEEtran}
\usepackage{latexsym}
\usepackage{graphicx}
\usepackage{amsfonts,amssymb,amsmath}
\def\BibTeX{{\rm B\kern-.05em{\sc i\kern-.025em b}\kern-.08em T\kern-.1667em\lower.7ex\hbox{E}\kern-.125emX}}

\usepackage{cite}
\usepackage{url}
\usepackage{soul}
\usepackage{multirow}
\usepackage{color}
\usepackage{hyperref}
\usepackage{breakurl}
\usepackage{epstopdf}
\usepackage{subfigure}
\usepackage{xcolor,pgfplots,tikz}

\soulregister\cite7
\soulregister\ref7

\begin{document}
	\title{Low-frequency Selection Switch based \\Cell-to-Cell Battery Voltage Equalizer with Reduced Switch Count}
	
	\author{
		\vskip 1em
		Shimul K. Dam, \emph{Student Member, IEEE}, and
		Vinod John, \emph{Senior Member, IEEE}
		
		\thanks{
			
			
			Shimul K. Dam and Vinod John are with the Department of Electrical Engineering, Indian Institute of Science, Bangalore, 560012, India (e-mail: shimuld@iisc.ac.in, vjohn@iisc.ac.in).

		}
		\thanks{This manuscript is an enhanced version of the work presented in 2019 IEEE Energy Conversion Congress and Exposition, Baltimore, USA, titled ``A selection switch based cell-to-cell battery voltage equalizer with reduced switch count''. The additional contents included in this manuscript are a) extensive literature survey with classification, b) discussion of different drive circuits for selection switches and advantages of low-frequency switch, c) equivalent MOSFET based implementation of selection switches, d) detailed explanation of the proposed cell voltage recovery compensation method and its benefits, e) a comparison with existing low-frequency cell-to-cell equalizers and a comparison with other types of equalizers, f) experimental results of equalization performance on eight-cell Li-ion stack, g) equalization performance under charging, discharging, and varying load conditions.}
	}
	
	\maketitle
	
	\begin{abstract}
		A selection switch based cell-to-cell voltage equalizer requires only one dual-port dc-dc converter shared by all the cells. A cell-to-cell voltage equalizer is proposed that utilizes a capacitively level-shifted Cuk converter and low-frequency cell selection switches. The absence of isolation transformer and diodes in the equalizer leads to high efficiency, and the use of low-frequency selection switches significantly reduces the cost of the drive circuits.  A low-frequency cell selection network is proposed using bipolar voltage buses, where the switch count is almost half, compared to the existing low-frequency cell-to-cell equalizers for the case of a large number of cells. A novel approach for cell voltage recovery compensation is proposed, which reduces the number of operations of the selection switches and the equalization time. The proposed equalizer is implemented with relays and verified with an 8-cell Li-ion stack. The developed prototype shows the efficiency of over 90\% and good voltage balancing performance during charging, discharging, and varying load conditions. Experimental results also show about one order of magnitude reduction in the number of relay switchings and a significant reduction in equalization time using the proposed voltage compensation.

	\end{abstract}
	
	\begin{IEEEkeywords}
		Cell-to-cell, cell balancing,  Cuk converter,  DPDT, low-frequency, selection switch, voltage equalizer.
	\end{IEEEkeywords}
	
	{}
	
	\definecolor{limegreen}{rgb}{0.2, 0.8, 0.2}
	\definecolor{forestgreen}{rgb}{0.13, 0.55, 0.13}
	\definecolor{greenhtml}{rgb}{0.0, 0.5, 0.0}
	
	\section{Introduction}
	
\begin{figure}[h!]
		\centering
		\includegraphics[width=7.0cm]{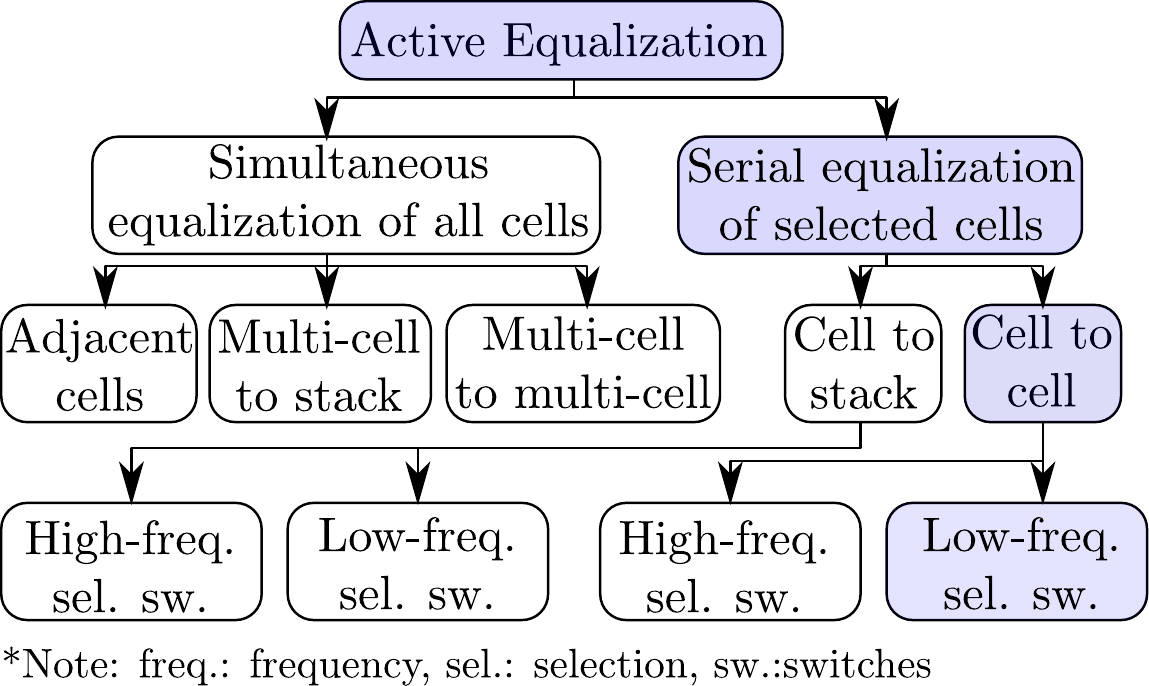}
		\caption{Classification of active voltage equalization methods.}
		\label{class}
	\end{figure}

	\IEEEPARstart{T}{he} cell voltages in a series-connected battery cell stack become unequal due to manufacturing tolerances, non-uniform aging, and unequal temperature distribution. The differences among the cell voltages increase with the number of charge-discharge cycles, leading to over-charge and over-discharge of some of the cells. A voltage equalizer is essential to improve the charge-capacity and the cycle life of the cell stack by avoiding such a situation.  
	The passive voltage equalizers, which are the simplest and cheapest equalizers, dissipate a significant amount of stored energy\cite{li_diss}.
	On the other hand, the active equalizers transfer charge from the over-charged cells to under-charged cells to equalize all the cell voltages\cite{14}. The operation of the active equalizers can be of two types: simultaneous equalization of all the cells and serial equalization of the selected cells, as shown in Fig.\,\ref{class}. 
	
	The first type of active equalizer uses multiple power converters or a multi-port power converter so that all the cells can take part in voltage equalization simultaneously. These equalizers can be classified into three categories: adjacent cell\cite{lee_int,lee_quasi,park_des,cassani_top,ye_zero,hua_cap}, multi-cell to stack\cite{einhorn,uno_double,chen,uno_single,hua_lifepo4,lim,hua_apwm,hua_rect,shang_mod,li}, and multi-cell to multi-cell\cite{ling,evzelman,yelaverthi,wang,shang_auto,ye_model,shang_delta,zeltser,ye_star,TPEL} equalizers, among which the multi-cell to multi-cell equalizers offer fastest voltage equalization. Each of these equalizers dedicates one converter port to each cell even though many of the cells in the cell stack may not require voltage equalization at a given point of time. Such dedicated connections result in a higher component count, under-utilization of converter components, and the requirement of many high-frequency isolated gate drivers. 
	
	The second type of active equalizer uses only one dual-port dc-dc converter shared by all the cells to reduce the component count and several selection switches for cell selection. 	
	When the number of cells increases, the dc-dc converter's component count remains the same, and only the number of selection switches increases. The charge transfer method in these equalizers is either cell-to-stack or cell-to-cell. The cell-to-stack equalizers \cite{imtiaz,nazi,lee_double,raber,    park_mod,kim_auto,kim_mod2,hannan,zhang_interleaved,lin} selects only one cell is at a time, and charge transfer takes place between that cell and the entire cell stack. On the other hand, cell-to-cell equalizers \cite{park_c2c,lee_cell_res,lee_cell_tr,yu  ,  xiong,pham,shang_cell} achieve a direct charge transfer from the most over-charged cell to the most under-charged cell. Thus, the cell-to-cell equalizers offer about twice the equalization speed, but they require twice as many selection switches. 
	
	As the number of selection switches is proportional to the number of cells, any simplification or cost-reduction of the selection switches and their drive circuits is valuable in case of a higher number of cells in the stack. As discussed in Section \ref{sec_drive_ckt} of this paper, a MOSFET, switched at a high-frequency, requires a complex gate-driver, which is often costlier than the MOSFET itself in case of low-power applications. Hence, the cell-to-stack equalizers in \cite{park_mod,kim_auto,kim_mod2,hannan,zhang_interleaved,lin} and cell-to-cell equalizers in\cite{xiong,pham,shang_cell} have been proposed with low-frequency selection switches so that simpler drive circuits can be used. This work aims to propose a Low-Frequency Selection Switch based Cell-to-Cell (LFSSCC) voltage equalizer with high efficiency, simple implementation, and lower switch count. The advantages of the proposed equalizer compared to existing LFSSCC equalizers\cite{xiong,pham,shang_cell} are discussed below.

	The dual-port dc-dc converters in \cite{xiong,pham} use diodes with significant conduction loss and transformer with higher high-frequency losses and size, leading to lower conversion efficiency and increased equalizer size. The dc-dc converter in \cite{shang_cell} achieves high-efficiency by achieving soft-switched operation with a higher component count. Thus, the dc-dc converters in the existing LFSSCC equalizers suffer from lower efficiency or higher circuit complexity. This work uses a capacitively level-shifted bidirectional Cuk converter, originally proposed for a multi-cell to multi-cell topology in \cite{ling}. The equalizer in \cite{ling} uses one closed-loop controlled Cuk converter for each cell, leading to a high component count and control requirements. On the other hand, this work uses only one Cuk converter for a large number of cells, reducing circuit complexity and control effort significantly. Thus, this converter is more suitable for a cell-to-cell equalizer and offers simpler implementation and high efficiency, as no transformer or diode is required in the conduction path. Section \ref{topology} discusses the converter topology.
	
	The LFSSCC equalizers\cite{xiong,pham,shang_cell} can use either low frequency switched MOSFETs or relays as selection switches. The selection switch networks in these equalizers require $2n$ Double Pole Double Throw (DPDT) switches for $n$ series-connected cells. These switches can be implemented with either $8n$ low-frequency MOSFETs or $2n$ DPDT relays. This work proposes a low-frequency selection switch network with $(n+2)$ DPDT and $2$ SPST switches, leading to a significant reduction in the number of switches and drive circuits.
	
	The voltage drop in cell impedance causes cell voltage recovery after equalization current is stopped. This voltage recovery leads to error in detecting end-of-equalization and a large number of switching of the selection switches. For a low-frequency selection switch based equalizer, a higher number of switching leads to longer equalization time and lower reliability. Different methods to reduce the number of selection switching are proposed based on voltage drop estimation\cite{hannan,pham} and self-learning fuzzy logic based method\cite{zhang_interleaved}. These methods require additional computation resources and cell charge-discharge characteristics. A simpler method with cell voltage recovery compensation, which does not require charge-discharge characteristics, is proposed to reduce the number of switchings of selection switches significantly.      
	
	This work in the enhanced version of the work published in \cite{ecce_volt_eq} with additional theoretical explanations and experimental results.
	Section \ref{selection} explains the proposed low-frequency cell selection network, and Section \ref{sec_volt_comp} discusses the proposed cell voltage recovery compensation. Sections \ref{sec_comp} and \ref{experiment} provides a comparison with existing equalizers and experimental validation of the proposed equalizer respectively.

	\section{Drive Circuits for Selection Switches}\label{sec_drive_ckt}	
	If a selection switch is operated at high frequency, it is implemented with MOSFETs. In contrast, a low-frequency selection switch is implemented with either MOSFETs or relay with significantly simpler and cheaper driver circuits.	
	
	\subsubsection{ High-Frequency MOSFET Drive Circuit} 
\begin{figure}[h!]
		\centering
		\begin{subfigure}[]{\includegraphics[width=2.1cm]{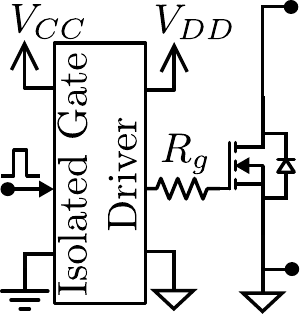}}
		\end{subfigure}
		\hspace{0.4cm}
		\begin{subfigure}[]{\includegraphics[width=2.1cm]{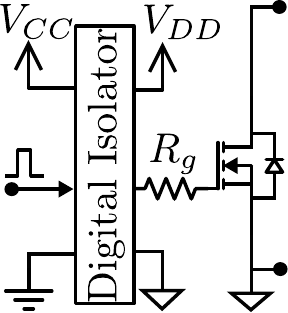}}
		\end{subfigure}
		\hspace{0.4cm}
		\begin{subfigure}[]{\includegraphics[width=2.3cm]{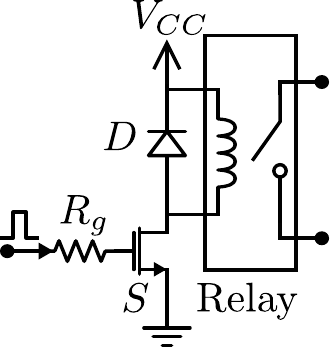}}
		\end{subfigure}		
		\caption{(a) High-frequency gate driver for MOSFET, (b) low-frequency gate driver for MOSFET, (c) driver for relay.}
		\label{drivers}
	\end{figure}

A high-frequency switched MOSFET for cell selection requires isolated gate driver IC, capable of providing a peak current of a few Ampere to achieve quick turn-on and turn-off, as shown in Fig.\,\ref{drivers}(a). Such a driver IC requires an isolated power supply and is often costlier than a low-power MOSFET. 
	
	\subsubsection{  Low-Frequency MOSFET Drive Circuit} Longer turn-on and turn-off times are acceptable for a low-frequency switched MOSFET. The gate drive resistor $R_g$ is in the order of a few kilo-Ohms, and the driver needs to supply only a few milli-Ampere peak current. Thus, a low-cost digital isolator can replace the costly gate driver IC, as shown in Fig.\,\ref{drivers}(b).
	
	\subsubsection{ Relay Drive Circuit} The driver circuit for a relay is simpler and cheaper than a MOSFET driver. Fig.\,\ref{drivers}(c) shows a relay driver, which requires a signal MOSFET $S$, a gate resistor $R_g$, and a free-whiling diode $D$. These components have to carry only a few tens of milli-Amperes of peak current, leading to a very low-cost implementation compared to high-frequency MOSFET drivers.
	
	Thus, the use of low-frequency selection switches reduces the complexity and cost of the equalizer significantly.

	\section{Dc-dc Converter Topology}\label{topology}
\begin{figure}[h!]
		\centering
		\includegraphics[width=5.5cm]{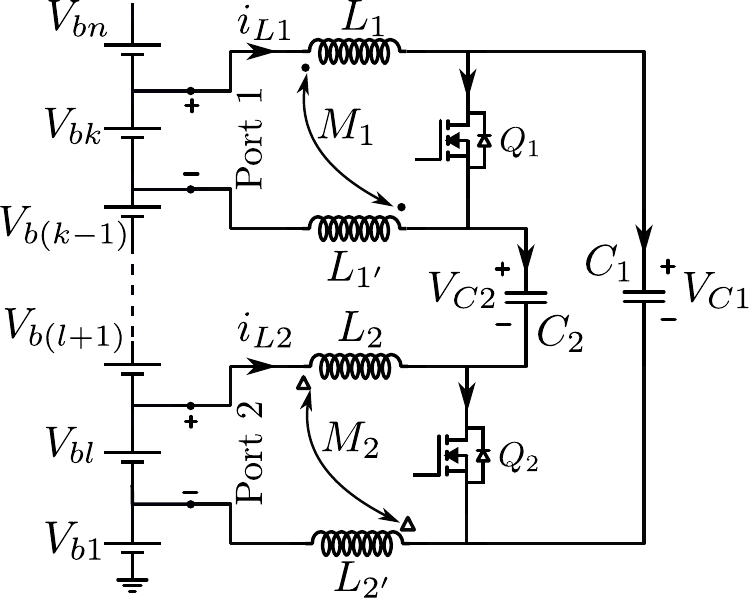}
		\caption{Schematic diagram of the capacitively level shifted Cuk converter connected to two cells of the stack.}
		\label{dc_dc_sch}
	\end{figure}

	The capacitively level-shifted Cuk converter allows variable voltage difference between the input and output ports' reference terminals without using any isolation transformer.  This converter is used here to transfer charge from the most over-charged cell to the most under-charged cell, as shown in Fig.\,\ref{dc_dc_sch}.
	
	This converter has an additional capacitor $C_2$ compared to a conventional bidirectional Cuk converter to block the voltage difference between the grounds of the two ports. Using KVL in Fig.\,\ref{dc_dc_sch} and assuming small voltage ripple, the voltage of the capacitors $C_1$ and $C_2$ are given by,
	\begin{eqnarray}
	\label{vc1}
	V_{C1}=\sum_{j=l}^{k}V_{bj} \qquad \text{and} \qquad V_{C2}=\sum_{j=l+1}^{k-1}V_{bj}
	\end{eqnarray}    
	Each inductor connected to the input or output port is split into two coupled inductors to make the circuit symmetric for reducing common mode oscillation. A more detailed discussion on the converter topology is provided in \cite{ecce_volt_eq}.
	A PI controller controls the cell current in one of the converter ports in closed-loop. 
	The converter is enabled when at least one cell voltage is out of the acceptable voltage range.
	
	For a predetermined tolerance voltage $V_{tol}$ and the average cell voltage $V_{avg}$, the $k^{th}$ cell voltage $V_{bk}$ is in the acceptable voltage range if it satisfies the following condition, 
	\begin{eqnarray}
	V_{avg}-V_{tol}\le V_{bk} \le V_{avg}+V_{tol} \quad \text{for all } k\in [1,n]
	\end{eqnarray}


	\section{Low-frequency Cell-to-cell Selection Network}\label{selection}
\begin{figure}[h!]
		\centering
		\begin{subfigure}[]{\includegraphics[width=3.8cm]{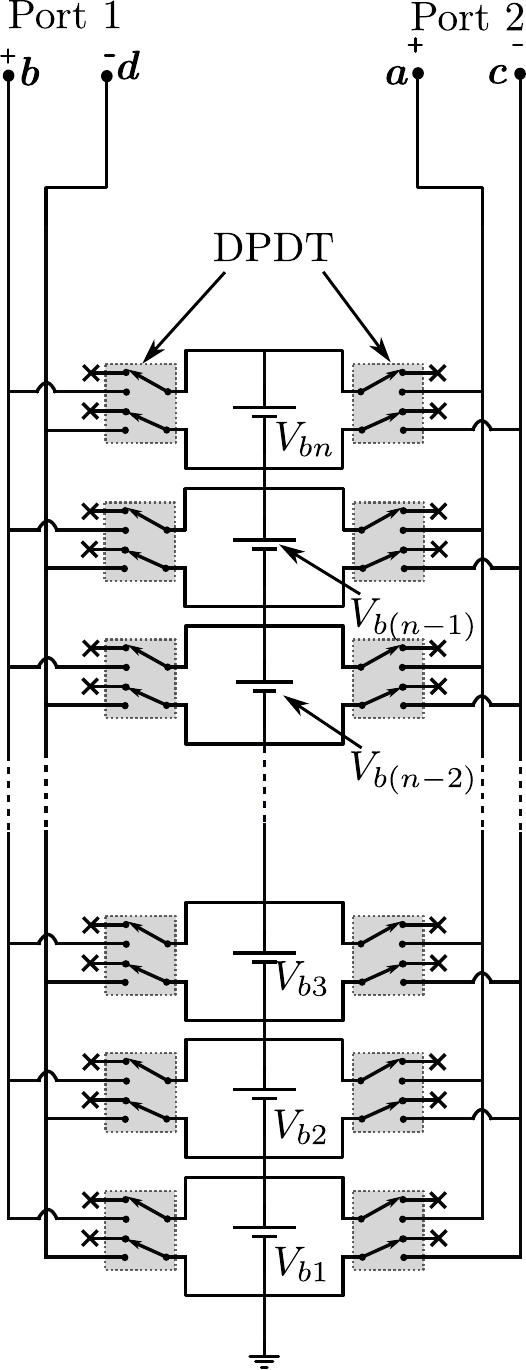}}
		\end{subfigure}\hspace{0.4cm}
		\begin{subfigure}[]{\includegraphics[width=3.3cm]{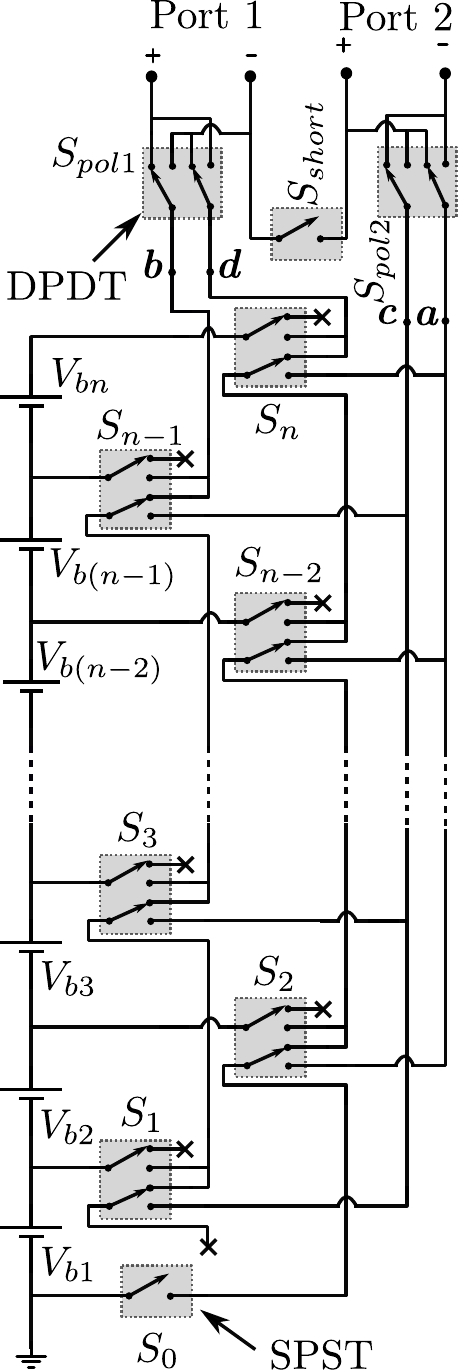}}
		\end{subfigure}
		\caption{Schematic diagram of the low-frequency cell selection network in (a) existing and (b) proposed cell-to-cell equalizers.}
		\label{dpdt}
	\end{figure}

	In an LFSSCC equalizer topology, it is possible to connect the dc-dc converter between any two cells at a time to achieve power transfer between them.  The existing LFSSCC equalizers\cite{shang_cell,xiong,pham} require $2n$ DPDT switches to implement such a cell selection network for $n$ series-connected cells, as shown in Fig.\,\ref{dpdt}(a), where the voltage rails \textbf{\textit{a}}, \textbf{\textit{b}}, \textbf{\textit{c}}, and \textbf{\textit{d}} are of fixed polarity. 	
	This work proposes a new low-frequency cell-to-cell selection network with bipolar voltage rails to reduce the selection switch count. 
	
	\begin{figure}[h!]
		\centering
		\begin{subfigure}[]{\includegraphics[width=3.2cm]{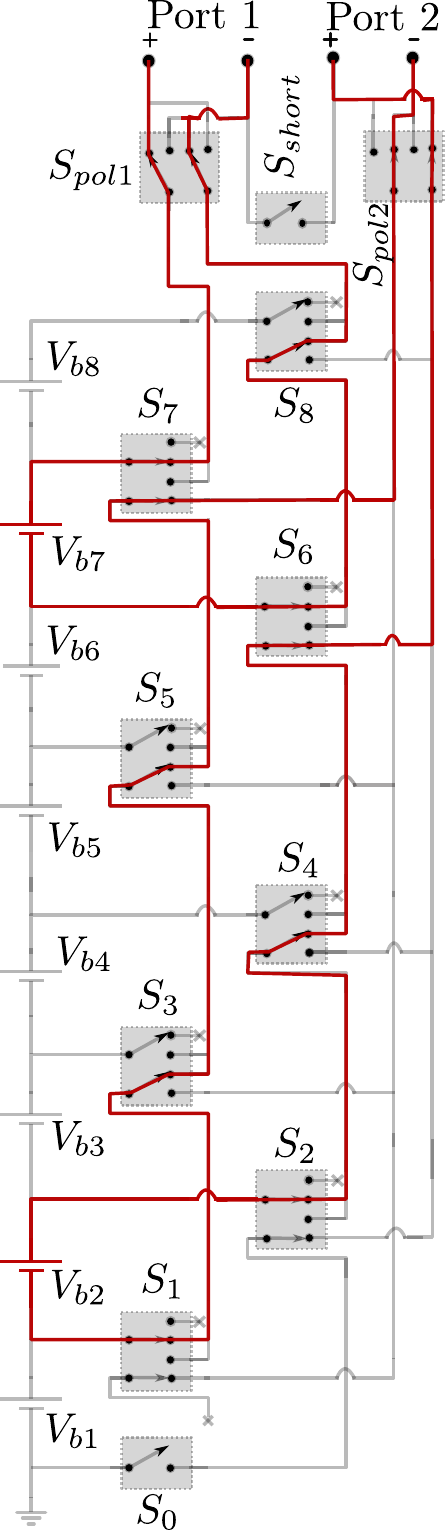}}
		\end{subfigure}\hspace{0.4cm}
		\begin{subfigure}[]{\includegraphics[width=3.2cm]{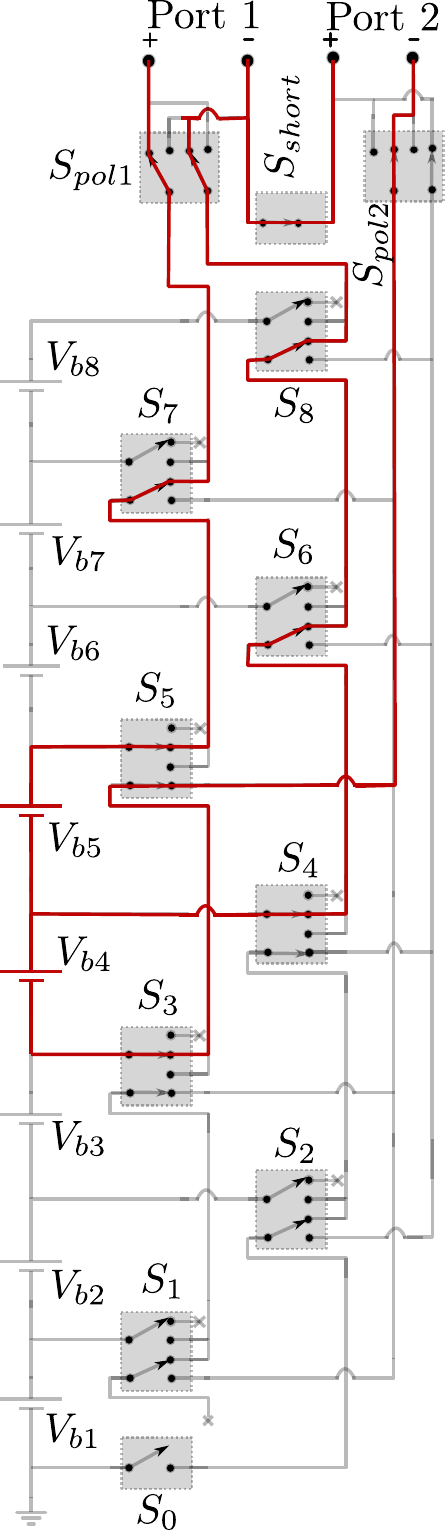}}
		\end{subfigure}
		\caption{Current flow paths between (a) the $2^{nd}$ and the $7^{th}$ cell, (b) the $4^{th}$ and the $5^{th}$ cell in the proposed low-frequency selection network.}
		\label{sel_net_8bat}
	\end{figure}

	\subsection{Proposed Cell Selection Network}
	Fig.\,\ref{dpdt}(b) shows the proposed low frequency cell-to-cell selection network, which uses One SPST switch and $n$ DPDT switches to select two cells at a time.
	One DPDT switch is used here to connect each common node between two adjacent cells to a voltage rail. However, each voltage rail is switched to a second rail if an upstream DPDT switch is turned on. Thus, each common node can effectively be connected to two voltage rails. In case the polarities of the voltage rails do not match with the converter port polarities, two DPDT relays $S_{pol1}$ and $S_{pol2}$ are used for polarity reversal. In this case, a polarity reversal is required for each pair of voltage rails when the corresponding cell is even-numbered.     
	In the case of two adjacent cells, the common node is connected to both converter ports by an SPST switch $S_{short}$. Thus, the proposed selection network requires ($n+2$) DPDT and $2$ SPST switches, and the selection switch count is reduced to almost half compared to the existing LFSSCC equalizers for a large number of cells.

	Lets consider that the selected cells are the $k^{th}$ and the $l^{th}$ cell, where $k>l$. Then, the $k^{th}$ cell is connected to port 1 and the $l^{th}$ cell is connected to port 2 with following steps,         
	\begin{itemize}
		\item Turn on $S_{pol1}$ if $k$ is even. 
		\item Turn on $S_{pol2}$ if $l$ is even. 
		\item Turn on $S_{short}$ if the $k^{th}$ and the $l^{th}$ cells are adjacent.
		\item Turn on $S_k$, $S_{k-1}$, $S_{l}$, and $S_{l-1}$.
	\end{itemize}
	%

	The selection network's operation is explained here for eight cells in two different equalization situations, as shown in Fig.\,\ref{sel_net_8bat}. The red lines indicate the current flow paths.
	
	\subsubsection{Two non-adjacent cells} Fig.\,\ref{sel_net_8bat}(a) shows the current flow path between two non-adjacent cells, cell 2 and cell 7. The switches  $S_{1}$, $S_{2}$, $S_{6}$, and $S_{7}$ are turned on to connect the cells to the converter ports.  The polarity reversal switch $S_{pol2}$ is turned on to ensure correct polarity connection.
	
	\subsubsection{Two adjacent cells} Fig.\,\ref{sel_net_8bat}(b) shows the current flow path for two adjacent cells, cell 4 and cell 5. The switches  $S_{3}$, $S_{4}$, $S_{5}$, and $S_{pol2}$ are turned on to connect cell 5 to port 1 and cell 4 to port 2 with the correct polarity. As the cell 4 and cell 5 are adjacent, their common node should be connected to the negative terminal of port 1 and positive terminal of port 2. This is achieved by turning on the shorting switch $S_{short}$.

	\subsection{Implementation of Selection Switches}
	The low-frequency selection switches can be implemented with MOSFETs or relays as discussed below,
	\subsubsection{DPDT switch for cell selection}
\begin{figure}[h!]
		\centering
		\begin{subfigure}[]{\includegraphics[width=2.7cm]{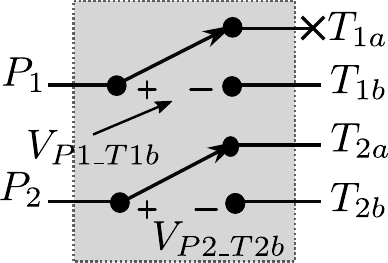}}
		\end{subfigure}\hspace{0.0cm}
		\begin{subfigure}[]{\includegraphics[width=2.7cm]{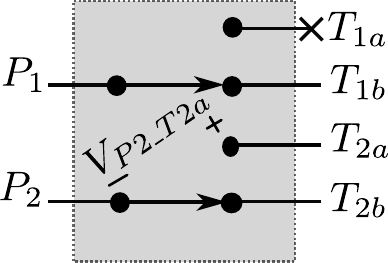}}
		\end{subfigure}     \hspace{0.0cm}     
		\begin{subfigure}[]{\includegraphics[width=2.7cm]{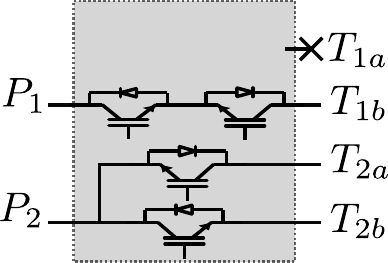}}
		\end{subfigure}
		\caption{DPDT switches for cell connection, $S_1$ to $S_n$: (a) off state, (b) on state, (c) a MOSFET based implementation.}
		\label{dpdt_mos}
	\end{figure}

	Fig.\,\ref{dpdt_mos}(a) and (b) show the off and on states of the DPDT switches $S_1$ to $S_n$. 	
	Each switch blocks the voltages $V_{P1\_T1b}$, $V_{P2\_T2b}$ in on-state, and $V_{P2\_T2a}$ in off-state. Fig.\,\ref{dpdt} and Fig.\,\ref{sel_net_8bat} show that $V_{P1\_T1b}$ is bipolar and $V_{P2\_T2b}$, $V_{P2\_T2a}$ are unipolar. Hence, the DPDT switch can be implemented with four MOSFETs, as shown in Fig.\,\ref{dpdt_mos}(c). Alternatively, it can also be implemented with a DPDT relay.
	\subsubsection{DPDT switch for polarity reversal}
\begin{figure}[h!]
		\centering
		\begin{subfigure}[]{\includegraphics[width=2.7cm]{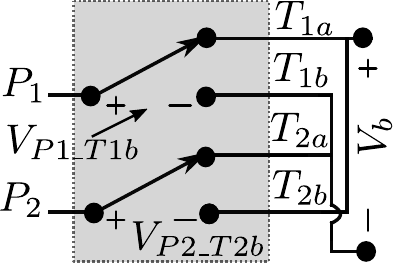}}
		\end{subfigure}\hspace{0.0cm}
		\begin{subfigure}[]{\includegraphics[width=2.7cm]{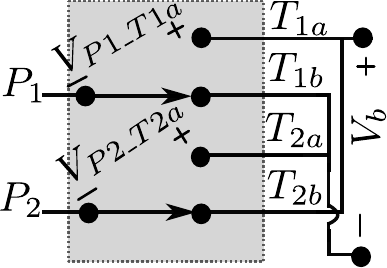}}
		\end{subfigure}     \hspace{0.0cm}     
		\begin{subfigure}[]{\includegraphics[width=2.7cm]{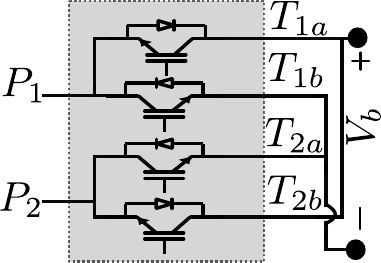}}
		\end{subfigure}
		\caption{DPDT switch for polarity reversal, $S_{pol1}$ and $S_{pol2}$: (a) off state, (b) on state, (c) a MOSFET based implementation.}
		\label{dpdt_pol_mos}
	\end{figure}

	Fig.\,\ref{dpdt_pol_mos}(a) and (b) show the off and on state of the polarity reversal DPDT switches $S_{pol1}$ and $S_{pol2}$. It can be observed that the voltage between each pole to each of its throw is positive or zero in off-state. The voltage between a throw and corresponding pole is positive or zero in on-state. The current in the DPDT switch is bi-directional. Hence, the DPDT switch can be implemented with a DPDT relay or four MOSFETs as shown in Fig.\,\ref{dpdt_pol_mos}(c).	
	
	\subsubsection{SPST switches}
	The SPST switches block unipolar voltages, and each of them can be implemented with an SPST relay or a MOSFET.

	Thus, the proposed selection switch network can be implemented with $(n+2)$ DPDT and $2$ SPST relays or with $(4n+10)$ MOSFETs.

	\section{Cell Voltage Recovery Compensation}\label{sec_volt_comp}
	When the discharging of a cell is stopped, the cell voltage $V_b$ recovers to a higher voltage after some time. Similarly, when the cell's charging is stopped, $V_b$ falls and settles to a lower voltage. This recovery in $V_b$ is often modeled as a voltage drop across the cell impedance. Ideally, when the cell's equalization is complete, $V_b$ should settle within the acceptable voltage band. However, if the cell's equalization is stopped when $V_b$ reaches within the acceptable voltage band, $V_b$ settles outside of the band due to voltage recovery. Thus, further rounds of equalization for the same cell are necessary, resulting in a higher number of switching of the selection switches.

	The control algorithm for a low-frequency cell selection network provides a small amount of time-gap, usually 10s to 30s for Li-ion cell, between two switching transitions for allowing the cell voltages to settle before selecting the next pair of cells for equalization. This time-gap helps to avoid high-frequency switching of the selection switches during transients in cell voltages. The measured cell voltages in low-frequency cell-to-stack\cite{kim_auto,lin} and cell-to-cell\cite{shang_cell} equalizers show the voltage recovery effect and resulting high number of charge or discharge rounds of each cell. Due to slow switching transitions and time-gap between transitions, a higher number of switchings leads to a longer equalization time. 
	
	Several attempts have been made to consider the effect of the cell voltage recovery within the equalization algorithm to avoid a higher number of switchings of selection switches. Estimation of the impedance drop\cite{hannan,pham} and self-learning fuzzy logic based method\cite{zhang_interleaved} have been employed for this purpose. The impedance drop estimation methods are computationally intensive. The self-learning fuzzy logic algorithm requires additional computation resources and prior knowledge of charge-discharge characteristics.
	A cell voltage recovery compensation is proposed here to reduce the number of switchings of the selection switches. The proposed method is computationally simpler than the impedance drop estimation methods and the fuzzy logic method. It does not require any prior knowledge of the charge-discharge characteristics of the cells.

	\begin{figure}[h!]
		\centering
		\includegraphics[width=8cm]{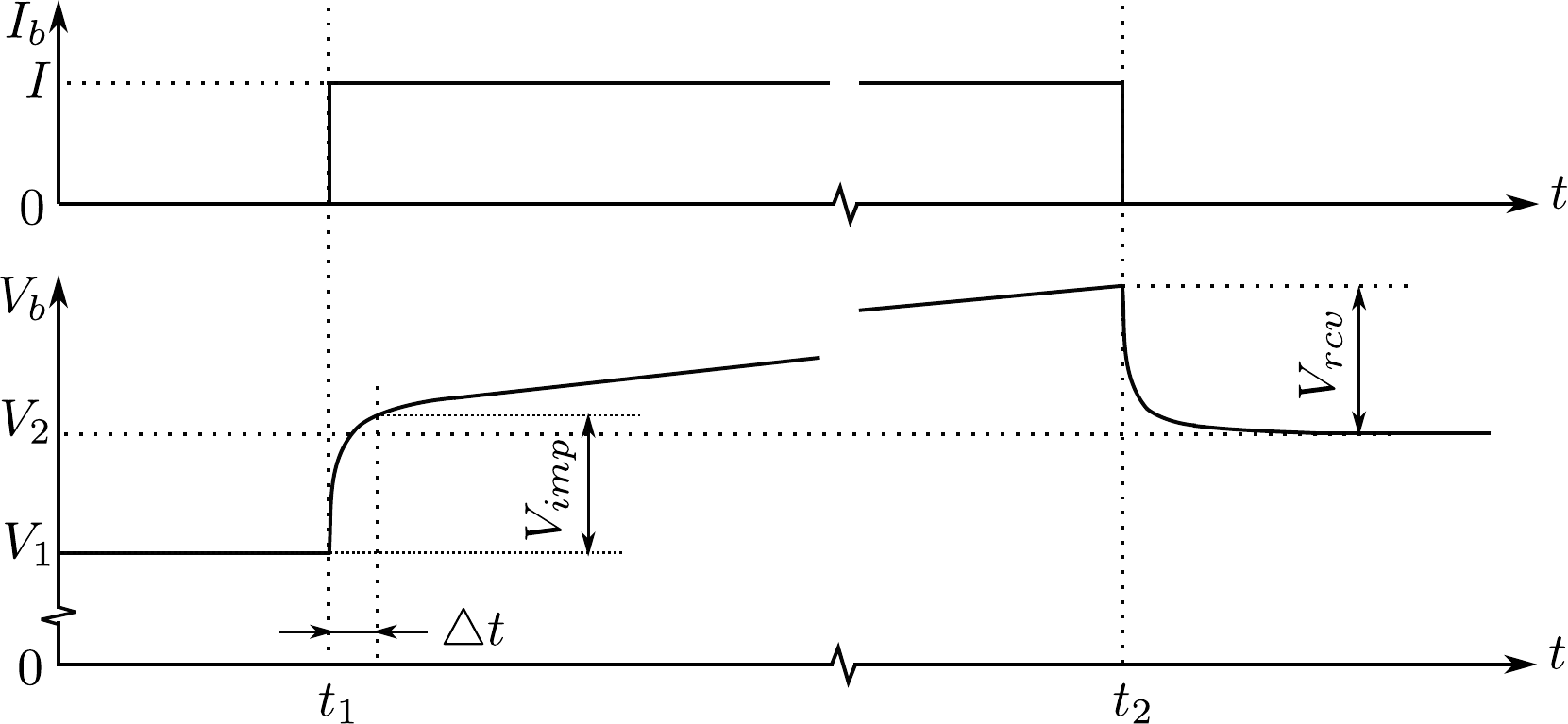}
		\caption{Cell voltage behavioral response when a step charging of the cell is initiated or terminated.}
		\label{volt_comp}
	\end{figure}

	The cell voltage recovery effect is explained here with an example of an under-charged cell. The equalization strategy is to charge the cell so that its voltage becomes equal to the average voltage of all the cells. The equalizer charges the under-charged cell with current $I$ from time $t_1$ to $t_2$. Fig.\,\ref{volt_comp} shows the cell voltage and current. 
	At the beginning of charging at $t_1$, $V_b$ rises from $V_1$ and mostly settles within a short time $\triangle t$.
	However, the cell voltage continues to increase slowly after time $(t_1+\triangle t)$ due to the charging of the cell. The charging is stopped at $t_2$ when the cell voltage is $(V_2+V_{rcv})$. The cell voltage $V_b$ then settles to voltage $V_2$. Thus, if the cell charging is stopped when $V_b$ reaches the average voltage $V_{avg}$, then $V_b$ settles to ($V_{avg}-V_{rcv}$).

	Hence, the equalizer should charge an under-charged cell until $V_b$ reaches $(V_{avg}+V_{rcv})$, so that $V_b$ settles to $V_{avg}$ after charging is stopped. Similarly, it should should discharge an over-charged cell until $V_b$ reduces to $(V_{avg}-V_{rcv})$. However, the estimation of $V_{rcv}$ with internal battery parameters requires significant computation efforts and often suffers from estimation error. In this work, the change in $V_b$ in time $\triangle t$ from the start of the charging is measured and stored in memory as $V_{imp}$. The voltage $V_{imp}$ is used as an estimate of $V_{rcv}$. However, $V_{rcv}$ is a function of cell condition and, hence, the use of $V_{imp}$ as an estimate of $V_{rcv}$ will have a lower error when the time duration $(t_2-t_1)$ is small.
	
	The cell voltage recovery compensation based algorithm in this work charges an under-charged cell till its voltage reaches $(V_{avg}+V_{imp})$ and discharges an over-charged cell till its voltage reduces to $(V_{avg}-V_{imp})$. 
	If the initial cell voltage $V_1$ is not close to the final cell voltage $V_2$, then $(t_2-t_1)$ is large and $V_{imp}$ is not a good estimate of $V_{rcv}$.  However, $V_b$ comes close to the acceptable voltage range after the first round of equalization and requires further charging or discharging for a shorter duration. In the second round of equalization, $(t_2-t_1)$ is small and $V_{imp}$ is a good approximation of $V_{rcv}$. Thus, $V_b$ settles within the acceptable voltage range within a few rounds of equalization.

	\begin{table*}[h]
		\centering
		\small
		\caption{Comparison of the proposed equalizer with the existing low-frequency selection switch cell-to-cell (LFSSCC) equalizers. \vspace{0.0cm}}
		\renewcommand{\arraystretch}{1.2}
		\renewcommand{\tabcolsep}{4pt}
		\begin{tabular}{|c|c|c|c|c|c|c|c|c|c|c|}
			\hline
			\multirow{3}{*}{\parbox{1.5cm}{\centering{Topology}}} &  \multirow{3}{*}{\parbox{2cm}{\centering{Type of selection switch}}} &  \multicolumn{8}{c|}{\centering{Number of components}}  & \multirow{3}{*}{\parbox{1.5cm}{\centering{Efficiency\\ (\%)}}} \\
			\cline{3-10}
			
			& & \multicolumn{3}{c|}{Selection switch network} & \multicolumn{5}{c|}{Dc-dc converter} & \\
			\cline{3-10}
			
			& & MOSFET & DPDT relay & SPST relay & MOSFET & Capacitor & Inductor & Transformer & Diode & \\
			\hline
			
			\multirow{2}{*}{Ref.\cite{xiong}}  & \parbox{1.8cm}{\centering{MOSFET}} & 8$n$ & 0 & 0 & \multirow{2}{*}{1} & \multirow{2}{*}{2} & \multirow{2}{*}{0} & \multirow{2}{*}{1} & \multirow{2}{*}{1} & \multirow{2}{*}{59.4}\\
			\cline{2-5} 
			& \parbox{1.8cm}{\centering{Relay}} & 0 & 2$n$ & 0 &  & & & & & \\
			\hline
			
			\multirow{2}{*}{Ref.\cite{pham}}  & \parbox{1.8cm}{\centering{MOSFET}} & 8$n$ & 0 & 0 & \multirow{2}{*}{2} & \multirow{2}{*}{2} & \multirow{2}{*}{0} & \multirow{2}{*}{1} & \multirow{2}{*}{2} & \multirow{2}{*}{\parbox{1.5cm}{\centering{85.3 to 89.5}}}\\
			\cline{2-5} 
			& \parbox{1.8cm}{\centering{Relay}} & 0 & 2$n$ & 0 & & & & & & \\
			\hline
			
			\multirow{2}{*}{Ref.\cite{shang_cell}}  & \parbox{1.8cm}{\centering{MOSFET}} & 8$n$ & 0 & 0 & \multirow{2}{*}{5} & \multirow{2}{*}{2} & \multirow{2}{*}{2} & \multirow{2}{*}{0} & \multirow{2}{*}{5} & \multirow{2}{*}{\parbox{1.5cm}{\centering{98.6 to 99.5}}}\\
			\cline{2-5} 
			& \parbox{1.8cm}{\centering{Relay}} & 0 & 2$n$ & 0 & & & & & & \\
			\hline
			
			\multirow{2}{*}{\parbox{1.5cm}{\centering{Proposed equalizer}}}  & \parbox{1.8cm}{\centering{MOSFET}} & 4$n$+10  & 0 & 0 & \multirow{2}{*}{2} & \multirow{2}{*}{2} & \multirow{2}{*}{2} & \multirow{2}{*}{0} & \multirow{2}{*}{0} & \multirow{2}{*}{\parbox{1.5cm}{\centering{90.1 to 92.9}}}\\
			\cline{2-5} 
			& \parbox{1.8cm}{\centering{Relay}} & 0 & $n$+2 & 2 & & & & & & \\
			\hline

			%
			%
		\end{tabular}
		\label{tab_comp_low_freq}
	\end{table*}

	\begin{table*}[h]
		\centering
		\small
		\caption{Comparison of the proposed low-frequency selection switch cell-to-cell (LFSSCC) equalizer with other types of equalizers. \vspace{0.0cm}}
		\renewcommand{\arraystretch}{1.2}
		\renewcommand{\tabcolsep}{2.5pt}
		\begin{tabular}{|c|c|c|c|c|c|c|c|c|c|c|c|c|c|}
			\hline
			\multirow{3}{*}{\parbox{1.2cm}{\centering{Topology}}} &  \multirow{3}{*}{\parbox{3.2cm}{\centering{Type of Equalizer}}} &  \multicolumn{7}{c|}{\centering{Number of components}} & \multicolumn{2}{c|}{\centering{Number of drivers}} & \multirow{3}{*}{\parbox{0.7cm}{\centering{Effici-ency (\%)}}} &  \multirow{3}{*}{\parbox{1.0cm}{\centering{Equaliza- tion speed}}} & \multirow{3}{*}{\parbox{1.5cm}{\centering{Voltage difference dependent}}} \\
			\cline{3-11}
			
			& & \multicolumn{2}{c|}{Semiconductor} & \multicolumn{2}{c|}{Relay} & \multicolumn{3}{c|}{Passives} &\multirow{2}{*}{\parbox{1.2cm}{\centering{High frequency}}} & \multirow{2}{*}{\parbox{1.2cm}{\centering{Low frequency}}} & & &\\
			\cline{3-9}
			
			& & MOSFET & Diode & DPDT & SPST & Cap. & Ind. & Trans. & & & & &  \\
			\hline
			
			Ref\cite{ye_zero}  &  \parbox{3cm}{\vspace{0.1 cm}\centering{Adjacent cell resonant-tank}\vspace{0.1 cm}} & 2$n$ & 0 & 0 & 0 & 2$n$-1 & $n$-1 & 0 & 2$n$ & 0 & 98.2 & Low  & Yes\\				
			\hline
			
			Ref\cite{li}  &  \parbox{3cm}{\vspace{0.1 cm}\centering{Multi-cell to stack multi-winding trans.}\vspace{0.1 cm}} & $n$+1 & 0 & 0 & 0 & $n$ & 0 & \parbox{0.7cm}{\centering{$n$+1 wind.}} & $n$+1 & 0 & 84.8 & Moderate  & Yes\\				
			\hline
			
			Ref\cite{ye_star}  &  \parbox{3cm}{\vspace{0.1 cm}\centering{Multi-cell to multi-cell switched capacitor}\vspace{0.1 cm}} & 2$n$ & 0 & 0 & 0 & 2$n$ & 0 & 0 & 2$n$ & 0 & - & Good  & Yes\\				
			\hline
			
			Ref\cite{wang}  &  \parbox{3cm}{\vspace{0.1 cm}\centering{Multi-cell to multi-cell dual-active bridge}\vspace{0.1 cm}} & 3$n$ & 0 & 0 & 0 & $n$ & 0 & $n$/2 & 3$n$ & 0 & 84.5 & Excellent  & No\\				
			\hline
			
			Ref\cite{hannan}  &  \parbox{3cm}{\vspace{0.1 cm}\centering{Cell-to-stack MOSFET based cell-selection}\vspace{0.1 cm}} & 4$n$+2 & 2 & 0 & 0 & 2 & 0 & 2 & 2 & 4$n$ & 92.0 & Moderate  &  No\\				
			\hline
			
			Ref\cite{lin}  &  \parbox{3cm}{\vspace{0.1 cm}\centering{Cell-to-stack Relay based cell-selection}\vspace{0.1 cm}} & 1 & 1 & 0 & 2$n$ & 2 & 2 & 0 & 1 & 2$n$ & - & Moderate  & No\\				
			\hline
			
			\multirow{2}{*}{Proposed}  &  \parbox{3.2cm}{\vspace{0.1 cm}\centering{MOSFET based cell-to-cell}\vspace{0.1 cm}} & 4$n$+12 & 0 & 0 & 0 & 2 & 2 & 0 & 2 & 4$n$+10 & \multirow{2}{*}{\parbox{0.7cm}{\centering{90.1-92.9}}} & \multirow{2}{*}{Good}  & \multirow{2}{*}{No}\\
			\cline{2-11}
			& \parbox{3.2cm}{\vspace{0.1 cm}\centering{Relay based cell-to-cell}\vspace{0.1 cm}} & 2 & 0 & $n$+2 & 2 & 2 & 2 & 0 & 2 & $n$+2 &  &  &\\				
			\hline
			
			\multicolumn{14}{l}{{$^*Note$: Cap.: Capacitor, Ind.: Inductor, Trans.: Transformer, wind.: winding, $n$ represents the number of cells.}}
			
		\end{tabular}
		\label{tab_comp_others}
	\end{table*}

	\section{Comparison of Cell-to-Cell Selection Networks}\label{sec_comp}
	A comparison of the component count and efficiency of the proposed equalizer with the existing LFSSCC equalizers, presented in Table \ref{tab_comp_low_freq}, shows that it can work with a lower number of selection switches compared to other LFSSCC equalizers. It can also be observed that the proposed equalizer achieves higher efficiency than \cite{xiong,pham} with similar complexity of the dc-dc converter, which is significantly simpler than \cite{shang_cell}. Thus, the proposed method achieves above 90\% efficiency with simple circuit implementation.
	
	Table \ref{tab_comp_others} compares the proposed LFSSCC equalizer with different types of existing equalizers. The followings can be observed from the comparison of component counts, driver requirements, efficiency, equalization speed, and voltage difference dependence,
	\begin{enumerate}
		\item Although the adjacent cell and multi-cell equalizers can work with a lower number of MOSFETs, they require a large number of passive components and high-frequency isolated gate drivers. 
		\item The equalization current in the adjacent cell and multi-cell equalizers, except \cite{wang}, is proportional to the cell voltage differences. Thus, these equalizers become less effective when the voltage difference is not large, especially in Li-ion cells, where even a large difference in SOC results in a small voltage difference. The work in \cite{wang} achieves voltage difference independence by controlling each of the cell currents simultaneously, leading to higher sensor and computation cost.
		\item The proposed equalizer offers similar component counts, driver requirements, and efficiencies compared to low-frequency cell-to-stack equalizers, but achieves twice as fast equalization.
		\item A relay-based implementation offers the lowest component count and driver requirements.
	\end{enumerate}

	\begin{figure}[h!]
		\centering
		
		\includegraphics[width=8.5cm]{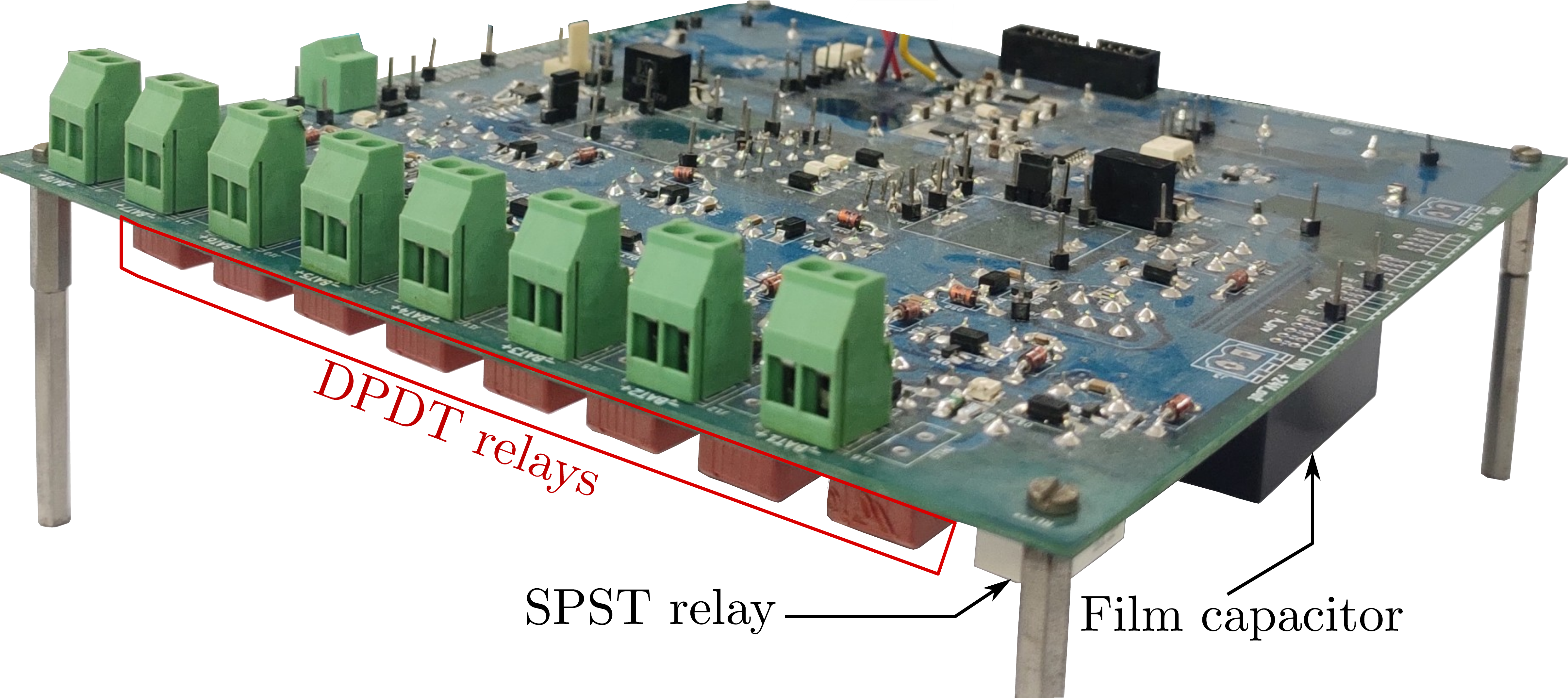}

		\caption{Image of the developed 8-cell equalizer prototype.}
		\label{setup_image}
	\end{figure}

	\begin{table}[h]
		\centering
		\small
		\caption{Circuit components and parameters of the proposed equalizer prototypes based on relays. \vspace{0.0cm}}
		\renewcommand{\arraystretch}{1.5}
		\renewcommand{\tabcolsep}{3pt}
		\begin{tabular}{|c|c|}
			\hline
			Component/parameter   &    Ratings/part no.    \\    \hline
			Switching frequency & $30$ kHz\\    \hline
			\parbox{5cm}{\centering{\vspace{0.0cm}  Inductances in port 1: $L_1$, $L_{1'}$, $M_1$}}  &  $62.5$ $\mu$H, $0.7$ A  \\    \hline
			\parbox{5cm}{\centering{\vspace{0.0cm}  Inductances in port 2: $L_2$, $L_{2'}$, $M_2$}}  &  $62.5$ $\mu$H, $0.7$ A  \\    \hline
			Capacitors: $C_1$, $C_2$ &  $50$ $\mu$F, $50$ V  \\    \hline
			MOSFETs: $Q_1$, $Q_2$   &   BSC009NE2LS5I \\    \hline
			SPST relay  &    OJE-SH-124LMH    \\    \hline
			DPDT relay  &    RT424024    \\    \hline
			Wait time for $V_{imp}$ measurement, $\triangle t$ & $20$ s\\ \hline

		\end{tabular}
		\label{tab_comp_ratings}
	\end{table}

	\section{Experimental Results}\label{experiment}
	An 8-cell prototype is developed to validate the proposed low-frequency selection switch cell-to-cell (LFSSCC) equalizer using DPDT relays as the selection switches. Table \ref{tab_comp_ratings} provides the ratings of the equalizer components, and Fig.\,\ref{setup_image} shows the image of the developed prototype. The prototype is tested with eight $3.6$ V, $2.6$ Ah Li-ion cells\cite{li_cell} to verify the converter operation, equalizer efficiency, and the control algorithm. The current of the cell connected to port $1$ of the converter is controlled to $0.5$ A, and the equalization algorithm decides its direction based on cell voltages.

	\subsection{Operation of Cuk Converter} 
\begin{figure}[h!]
		\centering
		\begin{subfigure}[]{\includegraphics[width=4cm]{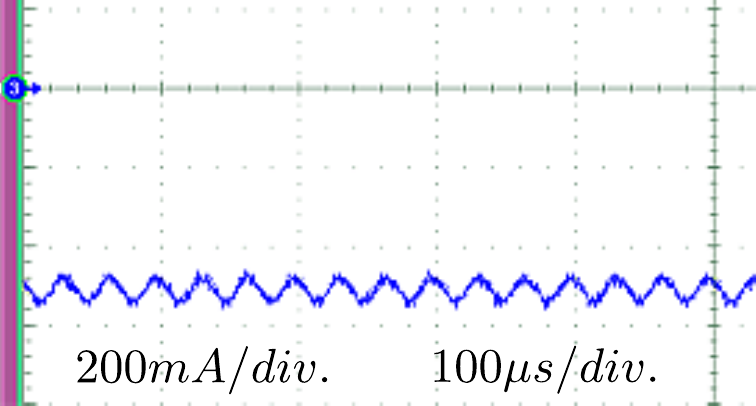}}
		\end{subfigure}\hspace{0.0cm}
		\begin{subfigure}[]{\includegraphics[width=4cm]{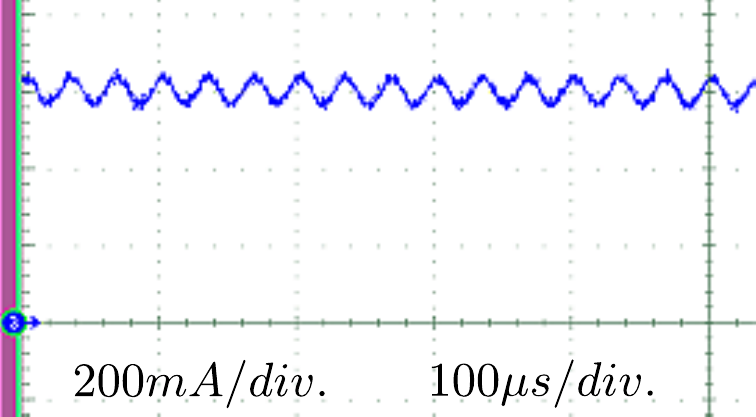}}
		\end{subfigure}
		\begin{subfigure}[]{\includegraphics[width=4.25cm]{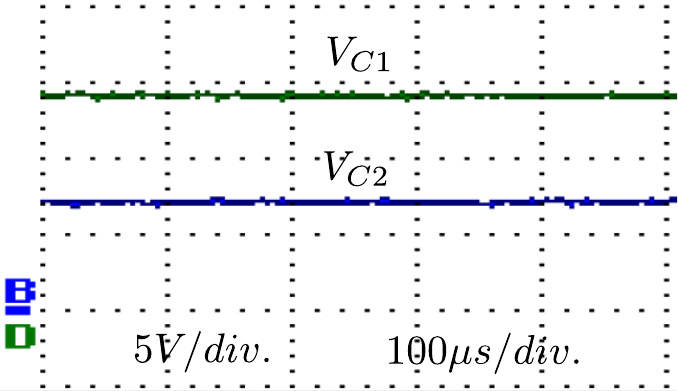}}
		\end{subfigure}\hspace{0.0cm}
		\begin{subfigure}[]{\includegraphics[width=4.25cm]{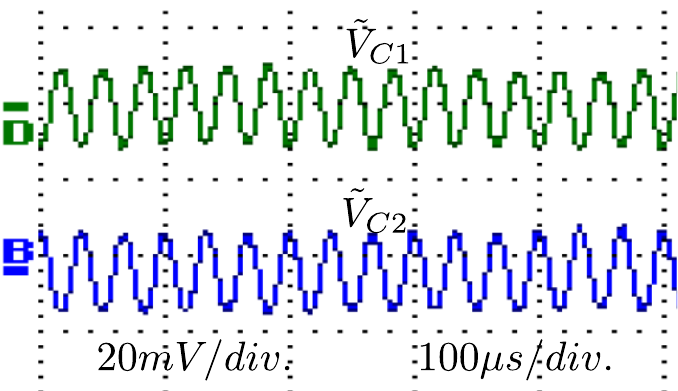}}
		\end{subfigure}
		
		\caption{Experimental waveforms of (a) port 1 current, (b) port 2 current, (c) capacitor voltages, and (d) capacitor voltage ripples of the Cuk converter.}
		\label{dpdt_wf}
	\end{figure}

	Fig.\ref{dpdt_wf} shows the measured current and voltage waveforms of the capacitively level-shifted Cuk converter in the prototype to verify the converter operation. The measured current waveforms show that the port 1 current is $0.5$ A, and the peak-to-peak ripple present in each of the port currents is $100$ mA. 
	Fig.\,\ref{dpdt_wf}(c) shows the measured capacitor voltages for the case when cell $1$ and cell $4$ are selected for equalization. It can be observed that the capacitor $C_1$ blocks the total voltage of $4$ cells, and $C_2$ blocks that total voltage of $2$ cells, as expected theoretically in (\ref{vc1}). The measured voltage ripples in $V_{C1}$ and $V_{C2}$ in Fig.\,\ref{dpdt_wf}(d) show peak-to-peak ripples of $20$ mV for both of the capacitors.

	%
	%
	
	\begin{figure}[h!]
		\centering
		
		\includegraphics[width=8.5cm]{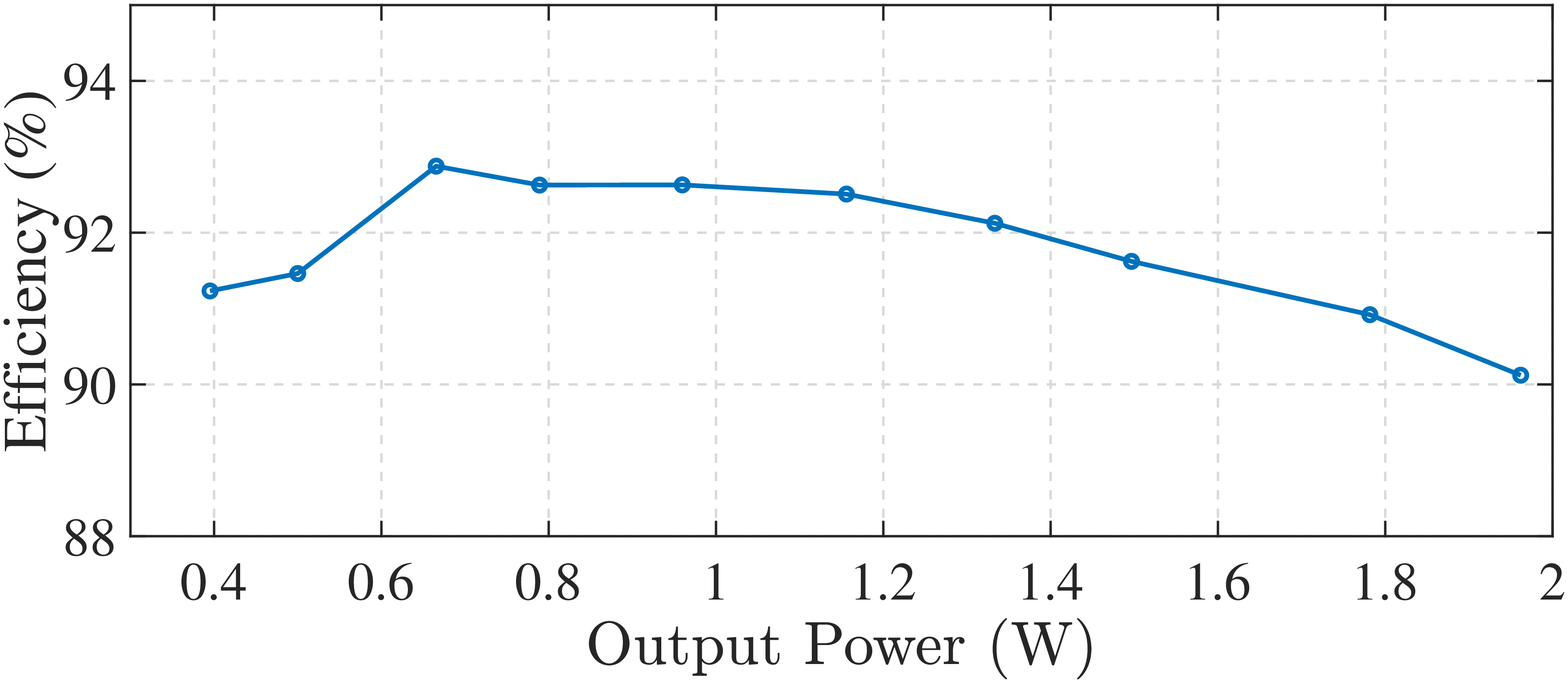}

		\caption{Plot of measured efficiency of the prototype with its output power.}
		\label{eff_plot}
	\end{figure}

	The efficiency of the prototype is measured for different output powers up to $2$ W and is plotted in Fig.\,\ref{eff_plot}. Fig.\,\ref{eff_plot} shows that the prototype has a maximum efficiency of $92.9\%$ and an efficiency of $90.1\%$ at the rated power.

	\begin{figure}[h!]
		\centering
		
		\begin{subfigure}[]{\includegraphics[width=8.8cm]{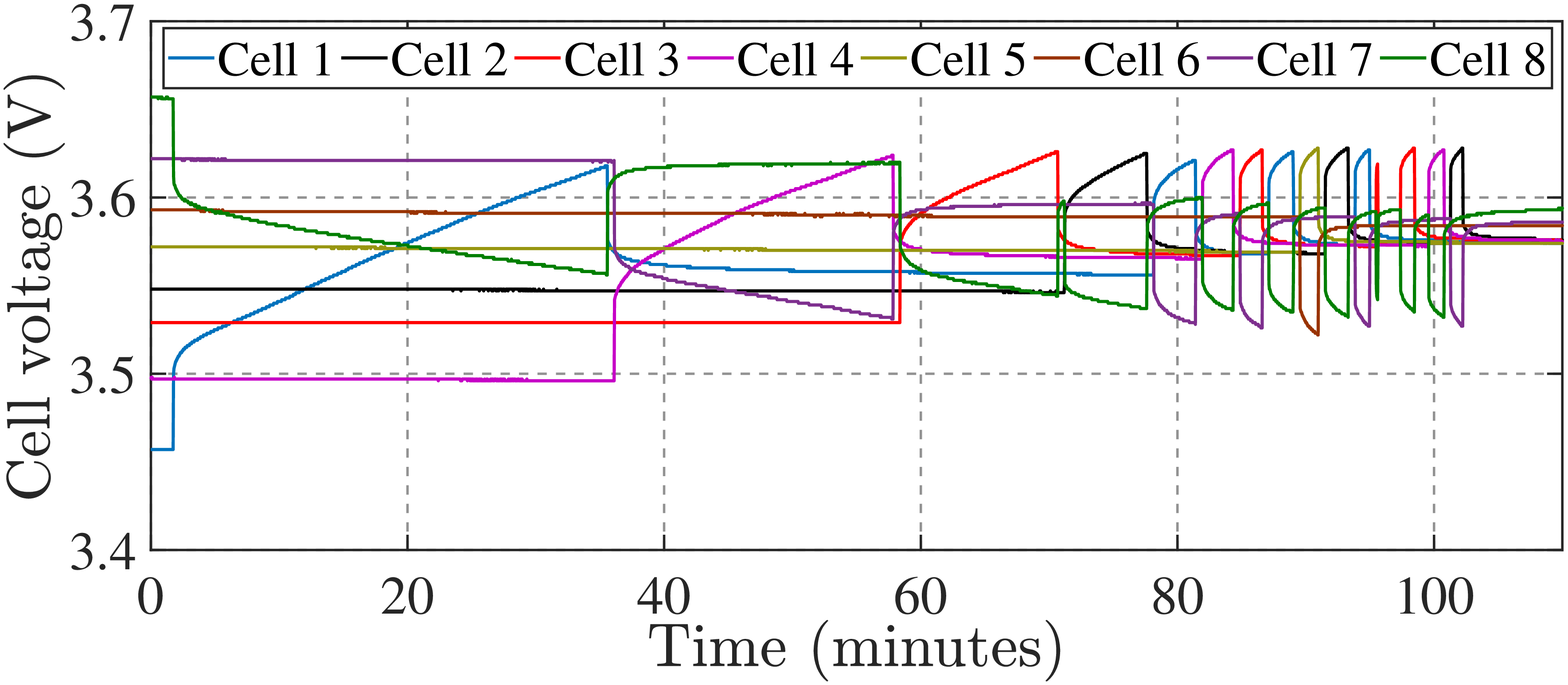}}
		\end{subfigure}\hspace{0.0cm}
		\begin{subfigure}[]{\includegraphics[width=8.8cm]{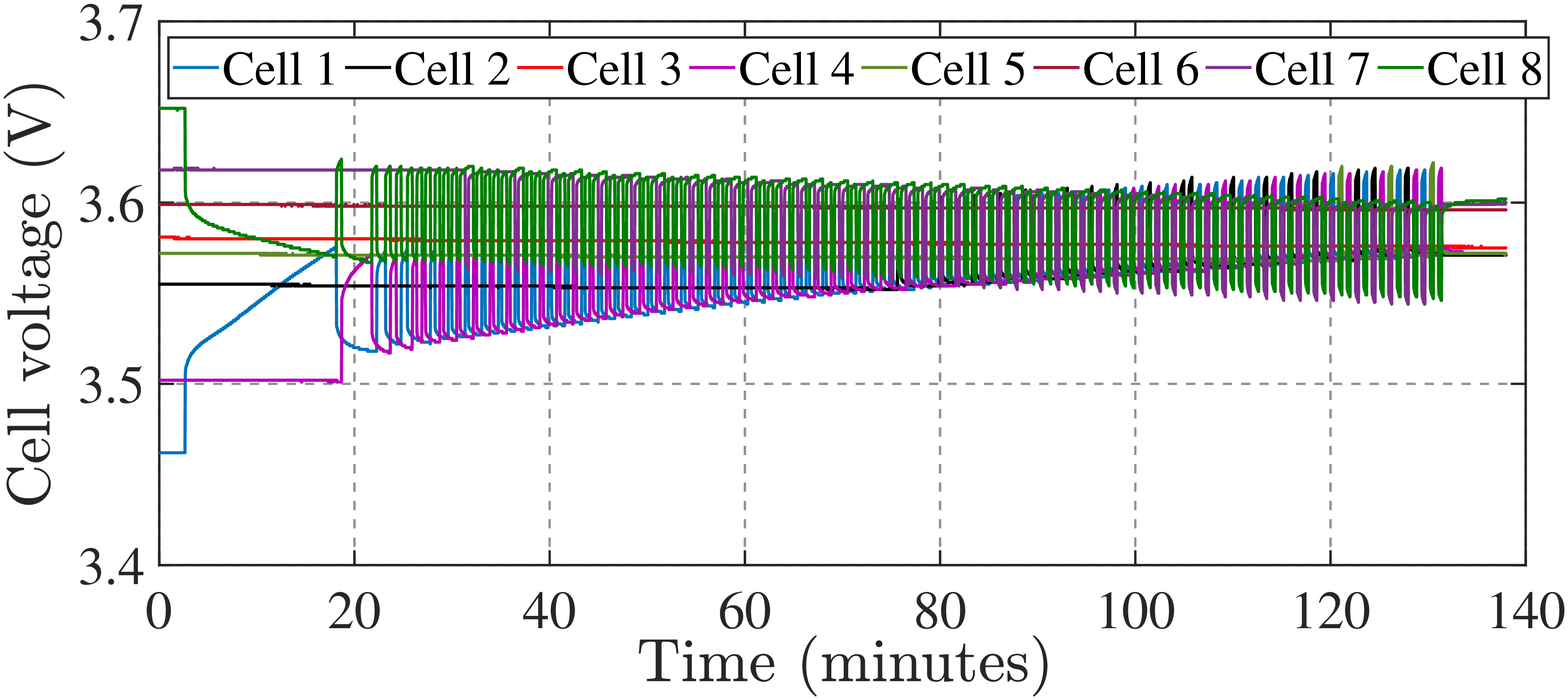}}
		\end{subfigure}
		
		\caption{Plot of measured cell voltages during voltage equalization (a) with, and (b) without cell voltage recovery compensation.}
		\label{spst_volt_conv}
	\end{figure}
	
	\subsection{Voltage Convergence Test}
	The performance of the equalizer and its control algorithm with the proposed cell voltage recovery compensation is tested on an 8-cell stack under rest condition, and Fig.\,\ref{spst_volt_conv}(a) shows the measured cell voltages. The maximum voltage difference is $200$ mV at the beginning. All the cell voltages converge within an acceptable voltage band of $20$ mV in $100$ minutes. 
	
	The same test is performed without the cell voltage recovery compensation to verify the necessity of this compensation, and Fig.\,\ref{spst_volt_conv}(b) shows the measured cell voltages. The initial cell voltages are close to those of the previous test for a proper comparison. The equalizer takes about $130$ minutes in this case to converge the cell voltages within a band of $30$ mV. Thus, the voltage convergence time is $30$\% longer when the cell voltage recovery compensation is not used. It can also be observed from Fig.\,\ref{spst_volt_conv}(a) and (b) that the number of switchings of the relays is significantly lower when cell voltage recovery compensation is employed. For example, the highest switched relay's number of switching transitions is reduced from $166$ to only $18$ using the compensation. The reduction of the number of switchings significantly improves the life and reliability of the relays.

	\begin{figure}[h!]
		\centering
		
		\includegraphics[width=8.8cm]{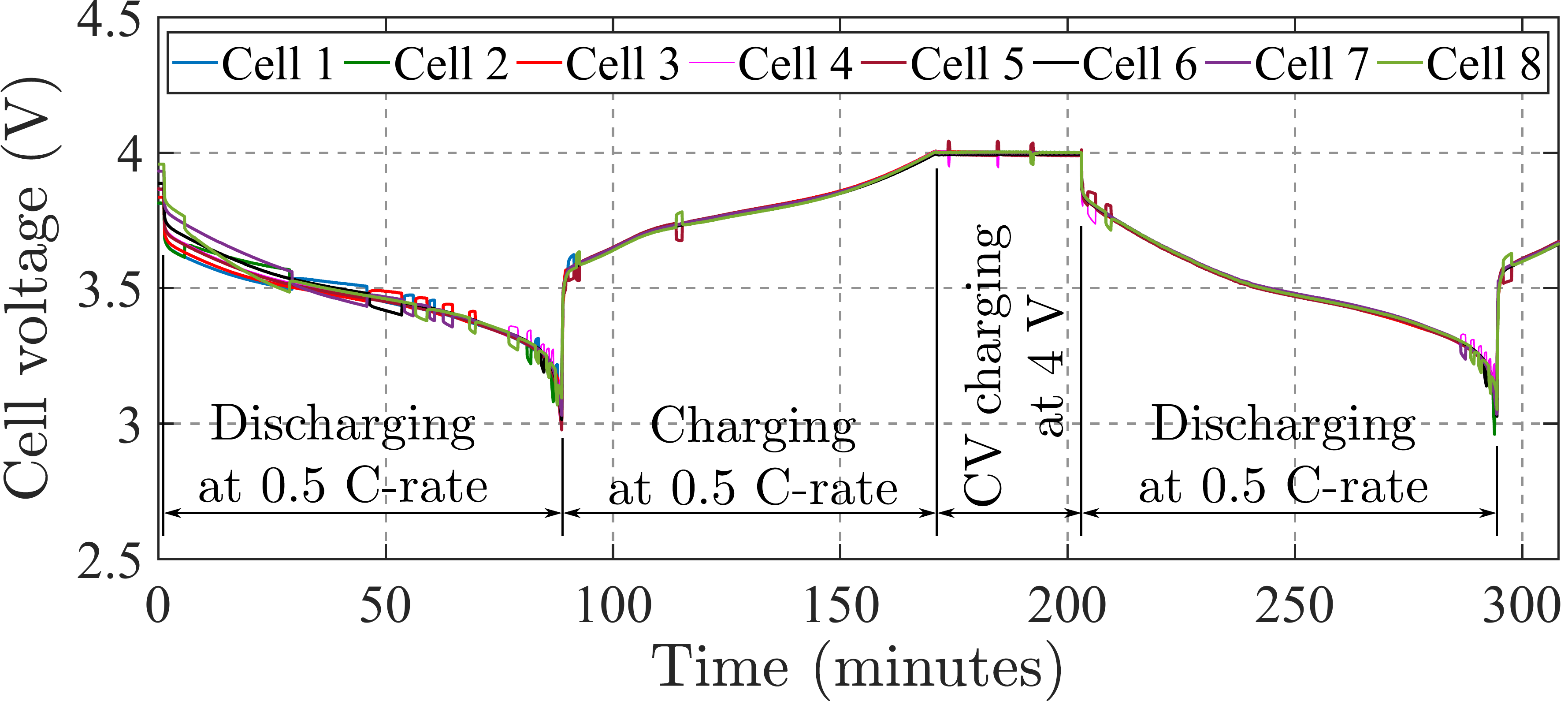}

		\caption{Plot of measured cell voltages over a charge-discharge cycle.}
		\label{volt_plot_ch_dis}
	\end{figure}
	
	\begin{figure}[h!]
		\centering
		
		\includegraphics[width=8.8cm]{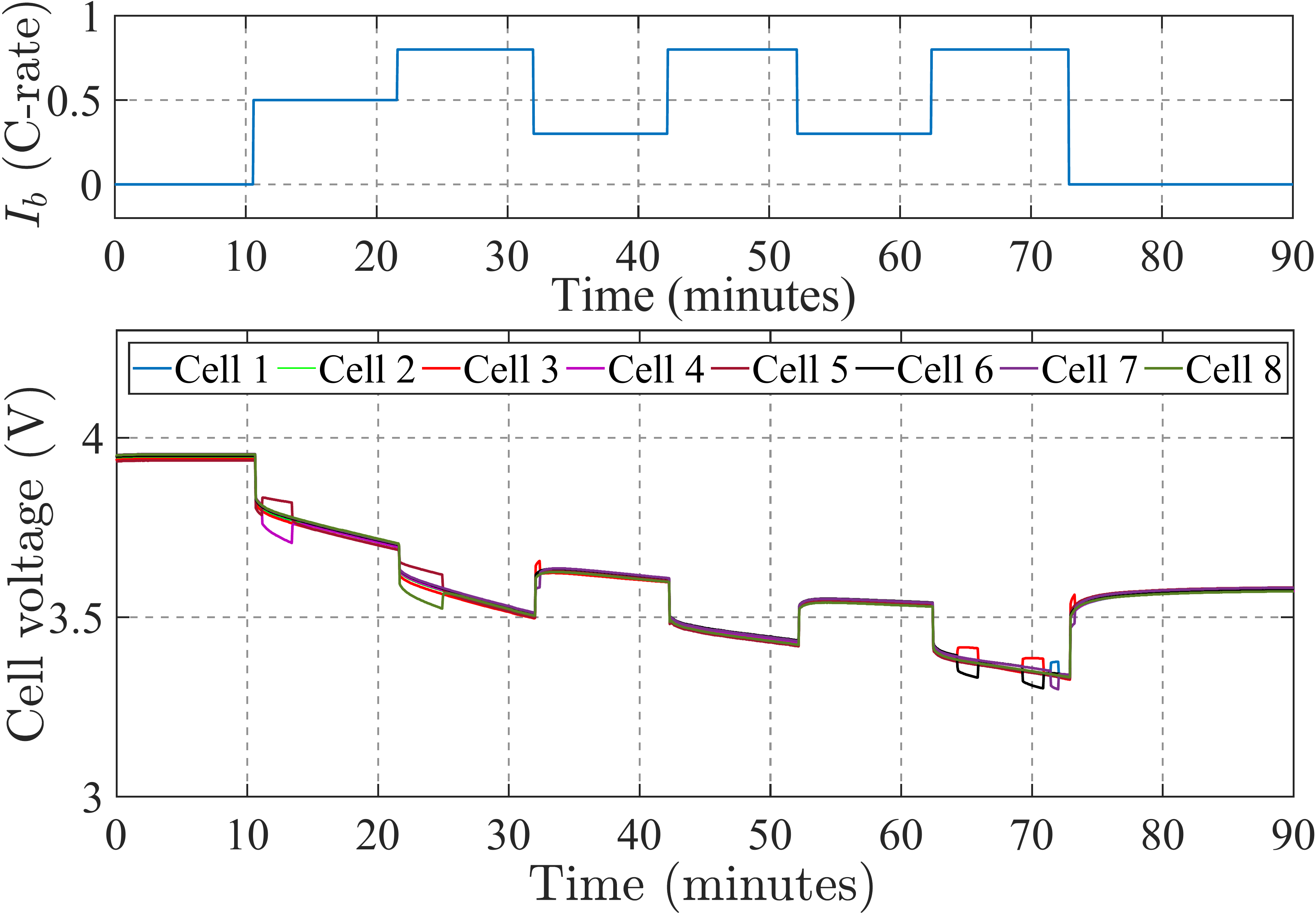}

		\caption{Plot of measured cell voltages with time during changes in load current $I_b$ of the cell stack.}
		\label{volt_plot_load}
	\end{figure}
	
	\subsection{Dynamic Performance of the Equalizer}
	The equalizer's performance is tested during the charge-discharge cycle of the eight-cell Li-ion stack, and Fig.\,\ref{volt_plot_ch_dis} shows the cell voltages. The cell voltages are initially unbalanced, with the maximum difference among them being $160$ mV. The cell stack is discharged from this condition at $0.5$ C-rate until the cell voltages reach $3$ V. The cell stack is then subjected to constant current (CC) charging at $0.5$ C-rate and constant voltage (CV) charging at $4$ V. Once, the CV charging is finished the stack is discharged once again at $0.5$ C-rate. It can be observed from the voltage plot that the equalizer equalizes the initially unbalanced cell voltages during the first discharge and maintains these voltages within $20$ mV voltage band. Once the cell voltages are equalized, the equalizer is disabled. However, on the course of charging and discharging, one or two cell voltages occasionally gets out of the $20$ mV voltage band the equalizer is enabled again, as observed near $115$, $170$, $180$, and $190$ minutes of Fig.\,\ref{volt_plot_ch_dis}. Fig.\,\ref{volt_plot_ch_dis} also shows that a very few switchings of the relays are required to maintain the cell voltages within the  $20$ mV band after the initial voltage difference is mitigated.  
	
	The performance of the equalizer is also verified under variable load conditions. Fig.\,\ref{volt_plot_load} shows the load current variation and the corresponding cell voltages.  It can be observed that a sudden change in current often produces unbalance in cell voltages, and the equalizer is activated to eliminate this unbalance. The equalizer is also activated when any cell voltage is out of the  $20$ mV voltage band during constant current discharging between two step-changes in current, as observed near $65$ and $70$ minutes of Fig.\,\ref{volt_plot_load}. Thus, the proposed equalizer meets performance requirements under different practical operating conditions.

	\section{Conclusion}
	A cell-to-cell voltage equalizer with low-frequency selection switches is proposed based on a capacitively level-shifted Cuk converter. The avoidance of an isolation transformer and diodes for the proposed equalizer's operation helps achieve high efficiency.    
	The use of low-frequency selection switches with simple drive circuits in the proposed equalizer leads to lower component count and cost.
	A low-frequency cell-to-cell selection network is proposed with bipolar voltage buses. This reduces the number of selection switches to almost half compared to the existing low-frequency selection networks for a large number of cells. 
	A comparison of the proposed equalizer with the existing LFSSCC equalizers shows its advantages in terms of switch count, drive circuit requirements, and efficiency. 
	An 8-cell prototype of the proposed equalizer is implemented with relays, and the operation of the capacitively level-shifted Cuk converter is experimentally verified. The developed prototype shows an efficiency above $90$\% over its entire power range and a peak efficiency of $92.9\%$. The equalizer is shown to successfully converge the voltages of eight Li-ion cells within a voltage band of $20$ mV from an initial voltage imbalance of $200$ mV. The prototype shows good voltage balancing performance under different conditions such as constant current charging, constant voltage charging, constant current discharge, and varying load.  
	A cell voltage recovery compensation scheme is proposed that estimates the voltage recovery due to cell impedance to reduce the number of switchings and the total equalization time. Experimental results show that the proposed compensation leads to the voltage convergence in a shorter time with about one order of magnitude reduction in the number of relay switchings. Thus, the proposed equalizer offers cell-to-cell voltage equalization with lower circuit complexity, component count, and good performance.

	
	\bibliographystyle{IEEEtranTIE}
	\bibliography{IEEEabrv,reference}\ 

\end{document}